\documentclass[final]{amsart}
\usepackage{times}
\usepackage{amsfonts,amscd,amsmath,amssymb,amsthm}

\numberwithin{equation}{section}

\renewcommand{\Im}{{\ensuremath{\mathrm{Im\,}}}}

\DeclareSymbolFont{SY}{U}{psy}{m}{n}
\DeclareMathSymbol{\emptyset}{\mathord}{SY}{'306}
\DeclareMathSymbol{\newtimes}{\mathbin}{SY}{'264}
\DeclareMathOperator*{\Bigtimes}{\newtimes}

\newcommand{\erfc}{\mathrm{erfc}}

\renewcommand{\kappa}{\varkappa}

\numberwithin{equation}{section}

 \DeclareMathOperator{\tr}{tr}
\DeclareMathOperator{\Ran}{Ran} \DeclareMathOperator{\Ker}{Ker}
 
\DeclareMathOperator{\Dom}{Dom} 
\DeclareMathOperator{\dist}{dist}

\newcommand{\R}{\mathbb{R}}

\newcommand{\C}{\mathbb{C}}
\newcommand{\Z}{\mathbb{Z}}
\newcommand{\N}{\mathbb{N}}

\newcommand{\1}{\mathbb{I}}

\newcommand{\fS}{\mathfrak{S}}

\newcommand{\fC}{\mathfrak{C}}

\newcommand{\fc}{\mathfrak{c}}
\newcommand{\fs}{\mathfrak{s}}

\newcommand{\cD}{{\mathcal D}}
\newcommand{\cE}{{\mathcal E}}

\newcommand{\cG}{{\mathcal G}}
\newcommand{\cH}{{\mathcal H}}
\newcommand{\cI}{{\mathcal I}}
\newcommand{\cJ}{{\mathcal J}}
\newcommand{\cK}{{\mathcal K}}
\newcommand{\cL}{{\mathcal L}}
\newcommand{\cM}{{\mathcal M}}

\newcommand{\cS}{{\mathcal S}}

\newcommand{\cW}{{\mathcal W}}

\newcommand{\diag}{\mathrm{diag}}

\newcommand{\sk}{\mathsf{k}}
\newcommand{\ii}{\mathrm{i}}
\newcommand{\e}{\mathrm{e}}

\newcommand{\bw}{\mathbf{w}}

\renewcommand{\epsilon}{\varepsilon}

\renewcommand{\det}{\mathrm{det}}

{\bf}{\it}
{\bf}{\it}
{\bf}{\it}
{\bf}{\it}

\newtheorem{theorem}{Theorem}[section]{\bf}{\it}
\newtheorem{proposition}[theorem]{Proposition}{\bf}{\it}
\newtheorem{corollary}[theorem]{Corollary}{\bf}{\it}
\newtheorem{example}[theorem]{Example}{\it}{\rm}
\newtheorem{lemma}[theorem]{Lemma}{\bf}{\it}
\newtheorem{remark}[theorem]{Remark}{\it}{\rm}
\newtheorem{definition}[theorem]{Definition}{\bf}{\it}

\title[Heat Kernels on Metric Graphs]{Heat Kernels on Metric Graphs and a Trace Formula}

\author[V. Kostrykin]{Vadim Kostrykin}
\address{Vadim Kostrykin\\ Institut f\"{u}r Mathematik, Technische Universit\"{a}t Clausthal,
Erz\-stra{\ss}e 1, D-38678 Clausthal-Zellerfeld, Germany}
\email{kostrykin@math.tu-clausthal.de, kostrykin@t-online.de}

\author[J. Potthoff]{J\"{u}rgen Potthoff}
\address{J\"{u}rgen Potthoff\\ Institut f\"{u}r Mathematik, Universit\"{a}t Mannheim, D-68131 Mannheim, Germany}
\email{potthoff@math.uni-mannheim.de}

\author[R. Schrader]{Robert Schrader}
\address{Robert Schrader\\ Institut f\"{u}r Theoretische Physik\\
Freie Universit\"{a}t Berlin, Arnim\-allee 14\\ D-14195 Berlin, Germany}
\email{schrader@physik.fu-berlin.de}

\dedicatory{Dedicated to Jean-Michel Combes on the occasion of his 65-th
birthday}

\copyrightinfo{2007}{V.~Kostrykin, J.~Potthoff, R.~Schrader}



\begin{document}

\begin{abstract}
We study heat semigroups generated by self-adjoint Laplace operators on
metric graphs characterized by the property that the local scattering
matrices associated with each vertex of the graph are independent from the
spectral parameter. For such operators we prove a representation for the
heat kernel as a sum over all walks with given initial and terminal edges.
Using this representation a trace formula for heat semigroups is proven.
Applications of the trace formula to inverse spectral and scattering
problems are also discussed.
\end{abstract}

\maketitle

\section{Introduction}\label{sec:intro}

Metric graphs or networks are one-dimensional piecewise linear spaces with
singularities at the vertices. Alternatively, a metric graph is a metric
space which can be written as a union of finitely many intervals, which are
either compact or $[0,\infty)$; any two of these intervals are either
disjoint or intersect only in one or both of their endpoints. It is natural
to call the metric graph compact if all its edges have finite length.

The increasing interest in the theory of differential operators on metric
graphs is motivated mainly by two reasons. The first reason is that such
operators arise in a variety of applications. We refer the reader to the
review \cite{Kuchment:0}, where a number of models arising in physics,
chemistry, and engineering is discussed. The second reason is purely
mathematical: It is intriguing to study the interrelation between the
spectra of these operators and topological or combinatorial properties of
the underlying graph. Similar interrelations are studied in spectral
geometry for differential operators on Riemannian manifolds (see,
e.g.~\cite{Chavel}, \cite{Gilkey}) and in spectral graph theory for
difference operators on combinatorial graphs (see, e.g.~\cite{Verdiere}).
Metric graphs take an intermediate position between manifolds and
combinatorial graphs.

In the present work we continue the study of heat semigroups on metric
graphs initiated in \cite{KS3}. There we provided sufficient conditions for
a self-adjoint Laplace operator to generate a contractive semigroup.
Moreover, we proved a criterion guaranteeing that this semigroup is
positivity preserving. For earlier work on heat semigroups generated by
Laplace operators on metric graphs and their application to spectral
analysis we refer to \cite{Angad-Guar}, \cite{Gaveau:1}, \cite{Gaveau:2},
\cite{Nicaise}, \cite{Nicaise:2}, \cite{Roth:1}, \cite{Roth:2}.

In this article we study heat semigroups generated by self-adjoint Laplace
operators which are characterized by the property that the local scattering
matrices associated with each vertex of the graph are independent of the
spectral parameter. All boundary conditions leading to such operators are
described in Proposition \ref{k-independent} below. In particular, Neumann,
Dirichlet, and the so called standard boundary conditions are in this class.

Our main technical tool to study heat semigroups on metric graphs are
\emph{walks} on edges of the graph, a concept developed in \cite{KS3},
\cite{KS4}. We will revisit this concept in Section \ref{sec:walks} below.
Furthermore, we will provide a representation for the heat kernel as a sum
over all walks with given initial and terminal edges. This representation
relates the topology of the graph to analytic properties of the heat
semigroup.

In Section \ref{sec:trac:form} we prove a trace formula for heat semigroups
on arbitrary (compact as well as noncompact) metric graphs, an analog of the
celebrated Selberg formula for differential operators on Riemannian
manifolds (see \cite{McKean}, \cite{Selberg} for the case of compact
manifolds and \cite{Hejhal}, \cite{Levay} for the noncompact case). A
discrete analog of the Selberg trace formula on $k$-regular trees is
discussed in \cite{Terras}.

The trace formula expresses the trace of the semigroup difference as the
sum over all \emph{cycles} on the graph, that is, equivalence classes of
closed walks. In the particular case of compact graphs and standard boundary
conditions our result recovers the well-known trace formula obtained by Roth
\cite{Roth:1}, \cite{Roth:2}. Related results can be found in \cite{KuNo},
\cite{Nicaise}, \cite{Nicaise:2}, \cite{Winn}. In the physical literature
trace formulas for Laplace operators on metric graphs have been discussed
in \cite{ACDMT}, \cite{Kottos}, \cite{Kottos:2}, \cite{Kottos:3}. Their
applications to quantum chaos and spectral statistics are reviewed in the
recent article \cite{GnuS}.

As an application of the trace formula, in Section \ref{sec:inv:probl} we
discuss inverse spectral and scattering problems. The inverse problems
considered here consist of determining the graph and its metric structure
(i.e.\ the lengths of its edges) from the spectrum of the Laplace operator
and the scattering phase (that is, half the phase of the determinant of the
scattering matrix), under the condition that the boundary conditions at all
vertices of the graph are supposed to be known. Another kind of the inverse
scattering problem, the reconstruction of the graph and the boundary
conditions from the scattering matrix, has been solved recently in
\cite{KS4}.

The results of Section \ref{sec:inv:probl} provide a mathematically
rigorous solution of the inverse scattering problem as proposed by Gutkin
and Smilansky in \cite{GS}. Also these result extend
the solution of the inverse spectral problem on compact graphs given by
Kurasov and Nowaszyk in \cite{KuNo} to more general boundary conditions.

\subsection*{Acknowledgments} It is a pleasure to thank the organizers of the conference
``Transport and Spectral Problems in Quantum Mechanics'' held at the
University of Cergy-Pontoise in September 2006 for a very interesting and
enjoyable meeting, both scientifically and socially. The authors would like
to thank M.~Karowski for helpful discussions.

\section{Background}\label{sec:back}

A finite graph is a 4-tuple $\cG=(V,\cI,\cE,\partial)$, where $V$ is a
finite set of \emph{vertices}, $\cI$ is a finite set of \emph{internal
edges}, $\cE$ is a finite set of \emph{external edges}. Elements in
$\cI\cup\cE$ are called \emph{edges}. The map $\partial$ assigns to each
internal edge $i\in\cI$ an ordered pair of (possibly equal) vertices
$\partial(i):=\{v_1,v_2\}$ and to each external edge $e\in\cE$ a single
vertex $v$. The vertices $v_1=:\partial^-(i)$ and $v_2=:\partial^+(i)$ are
called the \emph{initial} and \emph{terminal} vertex of the internal edge
$i$, respectively. The vertex $v=\partial(e)$ is the initial vertex of the
external edge $e$. If $\partial(i)=\{v,v\}$, that is,
$\partial^-(i)=\partial^+(i)$ then $i$ is called a \emph{tadpole}. A graph
is called \emph{compact} if $\cE=\emptyset$, otherwise it is
\emph{noncompact}.

Two vertices $v$ and $v^\prime$ are called \emph{adjacent} if there is an
internal edge $i\in\cI$ such that $v\in\partial(i)$ and
$v^\prime\in\partial(i)$. A vertex $v$ and the (internal or external) edge
$j\in\cI\cup\cE$ are \emph{incident} if $v\in\partial(j)$.

We do not require the map $\partial$ to be injective. In particular, any two
vertices are allowed to be adjacent to more than one internal edge and two
different external edges may be incident with the same vertex. If
$\partial$ is injective and $\partial^-(i)\neq\partial^+(i)$ for all
$i\in\cI$, the graph $\cG$ is called \emph{simple}.

The \emph{degree} $\deg(v)$ of the vertex $v$ is defined as
\begin{equation*}
\deg(v)=|\{e\in\cE\mid\partial(e)=v\}|+|\{i\in\cI\mid\partial^-(i)=v\}|+|\{i\in\cI\mid\partial^+(i)=v\}|,
\end{equation*}
that is, it is the number of (internal or external) edges incident with the
given vertex $v$ by which every tadpole is counted twice.

It is easy to extend the First Theorem of Graph Theory (see, e.g.\,
\cite{Diestel}) to the case of noncompact graphs:
\begin{equation}\label{1:thm:gt}
\sum_{v\in V} \deg(v) = |\cE| + 2 |\cI|.
\end{equation}

A vertex is called a \emph{boundary vertex} if it is incident with some
external edge. The set of all boundary vertices will be denoted by $\partial
V$. The vertices not in $\partial V$, that is in $V\setminus\partial V$ are
called \emph{internal vertices}.

The compact graph $\cG_{\mathrm{int}}=(V,\cI,\emptyset,\partial|_{\cI})$
will be called the \emph{interior} of the graph $\cG=(V,\cI,$ $
\cE,\partial)$. It is obtained from $\cG$ by eliminating the external
edges.

The \emph{star} $\cS(v)\subseteq \cE\cup\cI$ of the vertex $v\in V$ is the
set of the edges adjacent to $v$.

Throughout the whole work we will assume that the graph $\cG$ is connected,
that is, for any $v,v^\prime\in V$ there is an ordered sequence of vertices
$\{v_1=v, v_2,\ldots, v_{n-1}, v_n=v^\prime\}$ such that any two successive
vertices in this sequence are adjacent. In particular, this assumption
implies that any vertex of the graph $\cG$ has nonzero degree, i.e., for
any vertex there is at least one edge with which it is incident.

We will endow the graph with the following metric structure. Any internal
edge $i\in\cI$ will be associated with an interval $[0,a_i]$ with $a_i>0$
such that the initial vertex of $i$ corresponds to $x=0$ and the terminal
one to $x=a_i$. Any external edge $e\in\cE$ will be associated with a
semiline $[0,+\infty)$. We call the number $a_i$ the length of the internal
edge $i$. The set of lengths $\{a_i\}_{i\in\cI}$, which will also be treated
as an element of $\R^{|\cI|}$, will be denoted by $\underline{a}$. A
compact or noncompact graph $\cG$ endowed with a metric structure is called
a \emph{metric graph} and is written as $(\cG,\underline{a})$.

Given a finite graph $\cG=(V,\cI,\cE,\partial)$ with a metric structure
$\underline{a}=\{a_i\}_{i\in\cI}$ consider the Hilbert space
\begin{equation}\label{hilbert}
\cH\equiv\cH(\cE,\cI,\underline{a})=\cH_{\cE}\oplus\cH_{\cI},\qquad
\cH_{\cE}=\bigoplus_{e\in\cE}\cH_{e},\qquad
\cH_{\cI}=\bigoplus_{i\in\cI}\cH_{i},
\end{equation}
where $\cH_j=L^2(I_j)$ with
\begin{equation*}
I_j=\begin{cases} [0,a_j] & \text{if}\quad j\in\cI,\\  [0,\infty) &
\text{if}\quad j\in\cE.\end{cases}
\end{equation*}
Let $\overset{o}{I_j}$ be the interior of $I_j$, that is,
$\overset{o}{I_j}=(0,a_j)$ if $j\in\cI$ and $\overset{o}{I_j}=(0,\infty)$
if $j\in\cE$.

In the sequel the letters $x$ and $y$ will denote arbitrary elements of the
product set $\displaystyle\Bigtimes_{j\in\cE\cup\cI} I_j$.

By $\cD_j$ with $j\in\cE\cup\cI$ denote the set of all $\psi_j\in\cH_j$
such that $\psi_j(x)$ and its derivative $\psi^\prime_j(x)$ are absolutely
continuous and $\psi^{\prime\prime}_j(x)$ is square integrable. Let
$\cD_j^0$ denote the set of those elements $\psi_j\in\cD_j$ which satisfy
\begin{equation*}
\begin{matrix}
\psi_j(0)=0\\ \psi^\prime_j(0)=0
\end{matrix} \quad \text{for}\quad j\in\cE\qquad\text{and}\qquad
\begin{matrix}
\psi_j(0)=\psi_j(a_j)=0\\
\psi^\prime_j(0)=\psi^\prime_j(a_j)=0
\end{matrix}
\quad\text{for}\quad j\in\cI.
\end{equation*}
Let $\Delta^0$ be the differential operator
\begin{equation}\label{Delta:0}
\left(\Delta^0\psi\right)_j (x) = \frac{d^2}{dx^2} \psi_j(x),\qquad
j\in\cI\cup\cE
\end{equation}
with domain
\begin{equation*}
\cD^0=\bigoplus_{j\in\cE\cup\cI} \cD_j^0 \subset\cH.
\end{equation*}
It is straightforward to verify that $\Delta^0$ is a closed symmetric
operator with deficiency indices equal to $|\cE|+2|\cI|$.

We introduce an auxiliary finite-dimensional Hilbert space
\begin{equation}\label{K:def}
\cK\equiv\cK(\cE,\cI)=\cK_{\cE}\oplus\cK_{\cI}^{(-)}\oplus\cK_{\cI}^{(+)}
\end{equation}
with $\cK_{\cE}\cong\C^{|\cE|}$ and $\cK_{\cI}^{(\pm)}\cong\C^{|\cI|}$. Let
${}^d\cK$ denote the ``double'' of $\cK$, that is, ${}^d\cK=\cK\oplus\cK$.

For any $\displaystyle\psi\in\cD:=\bigoplus_{j\in\cE\cup\cI} \cD_j$ we set
\begin{equation}\label{lin1}
[\psi]:=\underline{\psi}\oplus \underline{\psi}^\prime\in{}^d\cK,
\end{equation}
with $\underline{\psi}$ and $\underline{\psi}^\prime$ defined by
\begin{equation}\label{lin1:add}
\underline{\psi} = \begin{pmatrix} \{\psi_e(0)\}_{e\in\cE} \\
                                   \{\psi_i(0)\}_{i\in\cI} \\
                                   \{\psi_i(a_i)\}_{i\in\cI} \\
                                     \end{pmatrix},\qquad
\underline{\psi}' = \begin{pmatrix} \{\psi_e'(0)\}_{e\in\cE} \\
                                   \{\psi_i'(0)\}_{i\in\cI} \\
                                   \{-\psi_i'(a_i)\}_{i\in\cI} \\
                                     \end{pmatrix}.
\end{equation}

Let $J$ be the canonical symplectic matrix on ${}^d\cK$,
\begin{equation}\label{J:canon}
J=\begin{pmatrix} 0& \1 \\ -\1 & 0
\end{pmatrix}
\end{equation}
with $\1$ being the identity operator on $\cK$. Consider the non-degenerate
Hermitian symplectic form
\begin{equation}\label{omega:canon}
\omega([\phi],[\psi]) := \langle[\phi], J[\psi]\rangle,
\end{equation}
where $\langle\cdot,\cdot\rangle$ denotes the scalar product in ${}^d
\cK\cong\C^{2(|\cE|+2|\cI|)}$.

A linear subspace $\cM$ of ${}^d\cK$ is called \emph{isotropic} if the form
$\omega$ vanishes identically on $\cM$. An isotropic subspace is called
\emph{maximal} if it is not a proper subspace of a larger isotropic
subspace. Every maximal isotropic subspace has complex dimension equal to
$|\cE|+2|\cI|$.

Let $A$ and $B$ be linear maps of $\cK$ onto itself. By $(A,B)$ we denote
the linear map from ${}^d\cK=\cK\oplus\cK$ to $\cK$ defined by the relation
\begin{equation*}
(A,B)\; (\chi_1\oplus \chi_2) := A\, \chi_1 + B\, \chi_2,
\end{equation*}
where $\chi_1,\chi_2\in\cK$. Set
\begin{equation}\label{M:def}
\cM(A,B) := \Ker\, (A,B).
\end{equation}

\begin{theorem}[\cite{KS1}]\label{thm:3.1}
A subspace $\cM\subset{}^d\cK$ is maximal isotropic if and only if there
exist linear maps $A,\,B:\; \cK\rightarrow\cK$ such that $\cM=\cM(A,B)$ and
\begin{equation}\label{abcond}
\begin{split}
\mathrm{(i)}\; & \;\text{the map $(A,B):\;{}^d\cK\rightarrow\cK$ has maximal
rank equal to
$|\cE|+2|\cI|$,}\qquad\\
\mathrm{(ii)}\; &\;\text{$AB^{\dagger}$ is self-adjoint,
    $AB^{\dagger}=BA^{\dagger}$.}
\end{split}
\end{equation}
\end{theorem}

Under the conditions \eqref{abcond} both $A\pm \ii\sk B$ are invertible for
all $\sk>0$.

\begin{definition}\label{def:equiv}
Two boundary conditions $(A,B)$ and $(A',B')$ satisfying \eqref{abcond} are
equivalent if the corresponding maximal isotropic subspaces coincide, that
is, $\cM(A,B)=\cM(A',B')$.
\end{definition}

The boundary conditions $(A,B)$ and $(A',B')$ satisfying \eqref{abcond} are
equivalent if and only if there is an invertible map $C:\,
\cK\rightarrow\cK$ such that $A'= CA$ and $B'=CB$ (see Proposition 3.6 in
\cite{KS4}).

By Lemma 3.3 in \cite{KS4}, a subspace $\cM(A,B)\subset{}^d\cK$ is maximal
isotropic if and only if
\begin{equation}\label{perp}
\cM(A,B)^{\perp} = \cM(B,-A).
\end{equation}
We mention also the equalities
\begin{equation*}
\begin{split}
\cM(A,B)^\perp & = \bigl[\Ker\,(A,B)\bigr]^\perp  = \Ran\, (A,B)^\dagger,\\
\cM(A,B) & = \Ran(-B,A)^\dagger.
\end{split}
\end{equation*}

There is an alternative parametrization of maximal isotropic subspaces of
${}^d\cK$ by unitary transformations in $\cK$ (see \cite{KS3} and
Proposition 3.6 in \cite{KS4}). A subspace $\cM(A,B)\subset {}^d\cK$ is
maximal isotropic if and only if for an arbitrary $\sk\in\R\setminus\{0\}$
the operator $A+\ii\sk B$ is invertible and
\begin{equation}\label{uuu:def}
\fS(\sk;A,B):=-(A+\ii\sk B)^{-1} (A-\ii\sk B)
\end{equation}
is unitary. Moreover, given any $\sk\in\R\setminus\{0\}$ the correspondence
between maximal isotropic subspaces $\cM\subset {}^d\cK$ and unitary
operators $\fS(\sk;A,B)\in\mathsf{U}(|\cE|+2|\cI|)$ on $\cK$ is one-to-one,
a result dating back to Bott \cite{Bott} and rediscovered in \cite{Arnold},
\cite{H}, and \cite{KS2}. Therefore, we will use the notation
$\mathfrak{S}(\sk;\cM)$ for $\fS(\sk;A,B)$ with $\cM(A,B)=\cM$.

Under the duality transformation $\cM\mapsto\cM^\perp$, as a direct
consequence of \eqref{perp} and \eqref{uuu:def}, the operators
\eqref{uuu:def} transform as follows (see Corollary 2.2 in \cite{KS1}):
\begin{equation}\label{sigma:perp}
\fS(\sk;\cM^\perp) = - \fS(\sk^{-1};\cM).
\end{equation}

There is a one-to-one correspondence between all self-adjoint extensions of
$\Delta^0$ and maximal isotropic subspaces of ${}^d\cK$ (see \cite{KS1},
\cite{KS4}). In explicit terms, any self-adjoint extension of $\Delta^0$ is
the differential operator defined by \eqref{Delta:0} with domain
\begin{equation}\label{thru}
\Dom(\Delta)=\{\psi\in\cD|\; [\psi]\in\cM\},
\end{equation}
where $\cM$ is a maximal isotropic subspace of ${}^d\cK$. Conversely, any
maximal isotropic subspace $\cM$ of ${}^d\cK$ defines through \eqref{thru}
a self-adjoint operator $\Delta(\cM, \underline{a})$. If $\cI=\emptyset$,
we will simply write $\Delta(\cM)$. In the sequel we will call the operator
$\Delta(\cM, \underline{a})$ a Laplace operator on the metric graph $(\cG,
\underline{a})$. {}From the discussion above it follows immediately that
any self-adjoint Laplace operator on $\cH$ equals
$\Delta(\cM,\underline{a})$ for some maximal isotropic subspace $\cM$.
Moreover, $\Delta(\cM,\underline{a})=\Delta(\cM^\prime,\underline{a})$ if
and only if $\cM=\cM^\prime$.

{}From Theorem \ref{thm:3.1} it follows that the domain of the Laplace
operator $\Delta(\cM,\underline{a})$ consists of functions $\psi\in\cD$
satisfying the boundary conditions
\begin{equation}\label{lin2}
 A\underline{\psi}+B\underline{\psi}^{\prime}=0,
\end{equation}
with $(A,B)$ subject to \eqref{M:def} and \eqref{abcond}. Here
$\underline{\psi}$ and $\underline{\psi}^{\prime}$ are defined by
\eqref{lin1:add}.

With respect to the orthogonal decomposition $\cK = \cK_{\cE}\oplus\cK_{\cI}^{(-)}\oplus\cK_{\cI}^{(+)}$
any element $\chi$ of
$\cK$ can be represented as a vector
\begin{equation}\label{elements}
\chi=\begin{pmatrix}\{\chi_e\}_{e\in\cE}\\ \{\chi^{(-)}_i\}_{i\in\cI}\\
\{\chi^{(+)}_i\}_{i\in\cI}\end{pmatrix}.
\end{equation}
Consider the orthogonal decomposition
\begin{equation}\label{K:ortho}
\cK = \bigoplus_{v\in V} \cL_{v}
\end{equation}
with $\cL_{v}$ the linear subspace of dimension $\deg(v)$ spanned by those
elements \eqref{elements} of $\cK$ which satisfy
\begin{equation}
\label{decomp}
\begin{split}
\chi_e=0 &\quad \text{if}\quad e\in \cE\quad\text{is not incident with the vertex}\quad v,\\
\chi^{(-)}_i=0 &\quad \text{if}\quad v\quad\text{is not an initial vertex of}\quad i\in \cI,\\
\chi^{(+)}_i=0 &\quad \text{if}\quad v\quad\text{is not a terminal vertex
of}\quad i\in \cI.
\end{split}
\end{equation}
Obviously, the subspaces $\cL_{v_1}$ and $\cL_{v_2}$ are orthogonal if
$v_1\neq v_2$.

Set ${^d}\cL_v:=\cL_v\oplus\cL_v\cong\C^{2\deg(v)}$. Obviously, each
$^d\cL_v$ inherits a symplectic structure from ${}^d\cK$ in a canonical way,
such that the orthogonal decomposition
\begin{equation*}
\bigoplus_{v\in V} {^d}\cL_v = {}^d\cK
\end{equation*}
holds.

\begin{definition}\label{propo}
Given the graph $\cG=\cG(V,\cI,\cE,\partial)$, boundary conditions $(A,B)$
satisfying \eqref{abcond} are called \emph{local on} $\cG$ if the maximal
isotropic subspace $\cM(A,B)$ of ${}^d\cK$ has an orthogonal symplectic
decomposition
\begin{equation}\label{propo:ortho}
\cM(A,B)=\bigoplus_{v\in V}\;\cM_v,
\end{equation}
with $\cM_v$ maximal isotropic subspaces of ${^d}\cL_v$. Otherwise the
boundary conditions are called \emph{non-local}.
\end{definition}

By Proposition 4.2 in \cite{KS4}, given the graph
$\cG=\cG(V,\cI,\cE,\partial)$, the boundary conditions $(A,B)$ satisfying
\eqref{abcond} are local on $\cG$ if and only if there is an invertible map
$C:\, \cK\rightarrow\cK$ and linear transformations $A(v)$ and $B(v)$ in
$\cL_{v}$ such that the simultaneous orthogonal decompositions
\begin{equation}\label{permut}
CA= \bigoplus_{v\in V} A(v)\quad \text{and}\quad CB= \bigoplus_{v\in V} B(v)
\end{equation}
are valid. {}From the equality $\cM(A,B)=\cM(CA, CB)$ it follows that the
subspaces $\cM_v$ in \eqref{propo:ortho} are equal to $\cM(A(v),B(v))$.

Boundary conditions $(A(v), B(v))$ induce local boundary conditions $(A,B)$
on the graph $\cG$ with
\begin{equation}\label{ab:decomp}
A= \bigoplus_{v\in V} A(v)\quad \text{and}\quad B= \bigoplus_{v\in V} B(v).
\end{equation}
{}From \eqref{permut} we get that
\begin{equation}\label{sdecomp}
\mathfrak{S}(\sk;A,B)=\mathfrak{S}(\sk;CA,CB)
 =\bigoplus_{v\in V}\mathfrak{S}(\sk;A(v),B(v))
\end{equation}
holds with respect to the orthogonal decomposition \eqref{K:ortho}.

The following proposition is taken from \cite{KS4}.

\begin{proposition}\label{k-independent}
Let $\cM=\cM(A,B)$ be a maximal isotropic subspace.
The following conditions are equivalent:
\begin{itemize}
\item[(i)]{$\mathfrak{S}(\sk;\cM)$ is $\sk$-independent,}
\item[(ii)]{$\mathfrak{S}(\sk;\cM)$ is self-adjoint for some $\sk>0$,}
\item[(iii)]{for some $\sk>0$ there is an orthogonal projection $P$ such that
$\mathfrak{S}(\sk;\cM)=\1-2P$,}
\item[(iv)]{$A B^\dagger =0$.}
\end{itemize}
\end{proposition}

Since this proposition will be crucial in what follows, we recall the

\begin{proof}
\textbf{(i) $\Leftrightarrow$ (ii).} Assume that $\mathfrak{S}(\sk;\cM)$ is
$\sk$-independent. Then, by \eqref{uuu:def}, for any eigenvector
$\chi\in\cK$ with eigenvalue $\lambda$ the equality
\begin{equation*}
(\lambda+1)A\chi + \ii\sk (\lambda-1) B\chi =0
\end{equation*}
holds for all $\sk>0$. Under the conditions \eqref{abcond} we have $\Ker A
\perp\Ker B$ (see Lemma 3.4 in \cite{KS4}). Hence, $\lambda\in\{-1,1\}$.
Thus, $\mathfrak{S}(\sk;\cM)$ is self-adjoint for all $\sk>0$.

Conversely, assume that $\mathfrak{S}(\sk;\cM)$ is self-adjoint for some
$\sk_0>0$. Due to the obvious equality
\begin{equation}\label{sinv}
\mathfrak{S}(\sk;\cM)=\bigl((\sk-\sk_{0})\mathfrak{S}(\sk_{0};\cM)+(\sk+\sk_{0})\bigr)^{-1}
\bigl((\sk+\sk_{0})\mathfrak{S}(\sk_{0};\cM)+(\sk-\sk_{0})\bigr),
\end{equation}
it is self-adjoint for all $\sk>0$. Let $\chi\in\cK$ be an arbitrary
eigenvector of $\mathfrak{S}(\sk_0;\cM)$ corresponding to the eigenvalue
$\lambda\in\{-1,1\}$. Observing that
\begin{equation*}
\frac{(\sk+\sk_0)\lambda+\sk-\sk_0}{(\sk-\sk_0)\lambda+\sk+\sk_0} = \lambda,
\end{equation*}
again by \eqref{sinv}, we conclude that  $\chi$ is an eigenvector of
$\mathfrak{S}(\sk;\cM)$ corresponding to the same eigenvalue $\lambda$ for all $\sk>0$.
Thus, $\mathfrak{S}(\sk;\cM)$ does not depend on $\sk>0$.

The equivalence \textbf{(ii) $\Leftrightarrow$ (iii)} is obvious.

The equivalence \textbf{(iv) $\Leftrightarrow$ (ii)} follows directly from
the identity
\begin{equation*}
\begin{split}
& \fS(\sk;\cM)-\fS(\sk;\cM)^\dagger\\ &\qquad = 2\ii\sk (A+\ii\sk
B)^{-1}\big[B(A^\dagger -\ii\sk B^\dagger) +(A+\ii\sk B)B^\dagger
\big](A^\dagger-\ii\sk B^\dagger)^{-1}\\ &\qquad = 4\ii\sk (A+\ii\sk
B)^{-1}A B^\dagger(A^\dagger-\ii\sk B^\dagger)^{-1}.
\end{split}
\end{equation*}
\end{proof}

We will write $\fS(\cM)$ instead of $\fS(\sk;\cM)$, whenever any of the
equivalent conditions of Proposition \ref{k-independent} is met.
Analogously we will drop the $\sk$-dependence in \eqref{sdecomp}:
\begin{equation*}
\mathfrak{S}({\cM})=\bigoplus_{v\in V}\mathfrak{S}(A(v),B(v))
=\bigoplus_{v\in V}\mathfrak{S}({\cM_v}).
\end{equation*}

{}From Proposition 3.5 in \cite{KS3} it follows that for any maximal
isotropic subspace $\cM$ satisfying any of the conditions of Proposition
\ref{k-independent}, the Laplace operator $-\Delta(\cM,\underline{a})$ is
nonnegative.

\begin{remark}
Assume that the maximal isotropic subspace $\cM\subset{}^d\cK$ satisfies any
of the conditions of Pro\-po\-sition \ref{k-independent}. By \eqref{perp}
the orthogonal maximal isotropic subspace $\cM^\perp\subset{}^d\cK$ then
also satisfies the conditions of Proposition \ref{k-independent}. {}From
\eqref{sigma:perp} it follows that $\fS(\cM^\perp)=-\fS(\cM)$.
\end{remark}

Obviously, Dirichlet $A=\1$, $B=0$ and Neumann $A=0$, $B=\1$ boundary
conditions satisfy the conditions of Proposition \ref{k-independent} with
$\fS(\1,0)=-\1$ and $\fS(0,\1)=\1$, respectively. We now provide two
important examples of boundary conditions satisfying the conditions
referred to in Proposition \ref{k-independent}.

\begin{example}[Standard boundary conditions]\label{3:ex:3}
Given a graph $\cG=\cG(V,\cI,\cE,\partial)$ for each vertex $v\in V$ with
$\deg(v)\geq 2$ define the boundary conditions $(A(v),B(v))$ the
$\deg(v)\times\deg(v)$ matrices
\begin{equation*}
\begin{aligned}
A(v)= \begin{pmatrix}
    1&-1&0&\ldots&&0&0\\
    0&1&-1&\ldots&&0 &0\\
    0&0&1&\ldots &&0 &0\\
    \vdots&\vdots&\vdots&&&\vdots&\vdots\\
    0&0&0&\ldots&&1&-1\\
    0&0&0&\ldots&&0&0
     \end{pmatrix},\quad
B(v)= \begin{pmatrix}
    0&0&0&\ldots&&0&0\\
    0&0&0&\ldots&&0&0\\
    0&0&0&\ldots&&0&0\\
    \vdots&\vdots&\vdots&&&\vdots&\vdots\\
    0&0&0&\ldots&&0&0\\
    1&1&1&\ldots&&1&1
\end{pmatrix}.
\end{aligned}
\end{equation*}
Clearly, $A(v)B(v)^\dagger=0$ and $(A(v),B(v))$ has maximal rank. The
corresponding unitary matrices \eqref{uuu:def} are given by
\begin{equation}\label{standard:ee}
[\mathfrak{S}(A(v),B(v))]_{e,e^\prime}= \frac{2}{\deg(v)} -
\delta_{e,e^\prime}
\end{equation}
with $\delta_{e,e^\prime}$ Kronecker symbol. If $\deg(v)=1$, we set
$A(v)=0$, $B(v)=1$ (Neumann boundary conditions) such that
\eqref{standard:ee} remains valid.

The local boundary conditions $(A,B)$ on the graph $\cG$ defined by
\eqref{ab:decomp} will be called \emph{standard} boundary conditions. We
use the notation $\cM_{\mathrm{st}}$ for the corresponding maximal
isotropic subspace.
\end{example}

\begin{remark}\label{3:rem:7}
Consider a graph with no internal lines
$\cG=(\{v\},\emptyset,\cE,\partial)$ and $|\cE|\geq 2$. By Proposition 2.1
in \cite{Exner:Turek}, the set of all isotropic subspaces satisfying any of
the equivalent conditions of Proposition \ref{k-independent}, contains
precisely four spaces, which correspond to the boundary conditions
invariant with respect to permutations of edges: $\cM(\1,0)$ (Dirichlet),
$\cM(0,\1)$ (Neumann), standard $\cM_{\mathrm{st}}$, and co-standard
$\cM_{\mathrm{st}}^\perp$.

Furthermore, by a result in \cite{KPS}, in the set of all isotropic
subspaces satisfying any of the equivalent conditions of Proposition
\ref{k-independent}, $\cM_{\mathrm{st}}$ is the only one with the property
that every function in the domain of $\Delta(\cM)$ is continuous at the
vertex $v$.
\end{remark}

\begin{example}[Magnetic perturbations of standard boundary conditions]\label{3:ex:mag}
If the maximal isotropic subspace $\cM(A,B)$ satisfies any of the equivalent
conditions of Proposition \ref{k-independent}, then for any unitary $U$ we
have
\begin{equation*}
AU (BU)^\dagger = A B^\dagger = 0.
\end{equation*}
Thus, the maximal isotropic subspace $\cM^U := \cM(AU, BU)$ also satisfies
the conditions of Proposition \ref{k-independent}. In particular, since
\begin{equation*}
\fS(\cM^U)= U^\dagger \fS(\cM) U,
\end{equation*}
we have the relation
\begin{equation}\label{spur}
\tr_{\cK} \fS(\cM^U) = \tr_{\cK} \fS(\cM).
\end{equation}

A special choice of unitary matrices $U$ corresponds to magnetic
perturbations of the Laplace operator $\Delta(\cM,\underline{a})$. By a
result in \cite{KS5} any magnetic perturbation of the Laplace operator
$\Delta(\cM,\underline{a})$ is unitarily equivalent to
$\Delta(\cM^U,\underline{a})$ with some $\displaystyle U=\bigoplus_{v\in V}
U_v$, where every $U_v$ is unitary and diagonal with respect to the
canonical basis in $\cL_v$,
\begin{equation*}
U_v = \diag\left(\{\e^{\ii\varphi_j(v)}\}_{j\in\cS(v)}\right).
\end{equation*}
In particular, any magnetic perturbation of standard boundary conditions
(see Example \ref{3:ex:3}) satisfies the conditions of Proposition
\ref{k-independent}.
\end{example}

\section{Heat Kernel and Walks on the Graph}\label{sec:walks}

\subsection{The Resolvent}

The structure of the underlying Hilbert space $\cH$ \eqref{hilbert} gives
naturally rise to the following definition of integral operators.

\begin{definition}
The operator $K$ on the Hilbert space $\cH$ is called \emph{integral
operator} if for all $j,j^\prime\in\cE\cup\cI$ there are measurable
functions $K_{j,j^\prime}(\cdot,\cdot)\, :\, I_j\times
I_{j\prime}\rightarrow \C$ with the following properties
\begin{itemize}
\item[(i)]{$K_{j,j^\prime}(x_j,\cdot)\varphi_{j^\prime}(\cdot)\in L^1(I_{j^\prime})$
for almost all $x_j\in I_j$,}
\item[(ii)]{$\psi=K\varphi$ with
\begin{equation}\label{kern}
\psi_j(x_j) = \sum_{j^\prime\in\cE\cup\cI} \int_{I_{j^\prime}}
K_{j,j^\prime}(x_j,y_{j^\prime}) \varphi_{j^\prime}(y_{j^\prime})
dy_{j^\prime}.
\end{equation}}
\end{itemize}
The $(|\cI|+|\cE|)\times(|\cI|+|\cE|)$ matrix-valued function $(x,y)\mapsto
K(x,y)$ with
\begin{equation*}
[K(x,y)]_{j,j^\prime} = K_{j,j^\prime}(x_j,y_{j^\prime})
\end{equation*}
is called the \emph{integral kernel} of the operator $K$.
\end{definition}

Below we will use the following shorthand notation for \eqref{kern}:
\begin{equation*}
\psi(x) = \int^{\cG} K(x,y) \varphi(y) dy.
\end{equation*}

We denote
\begin{equation}\label{U:def:neu}
R(\sk;\underline{a}) := \begin{pmatrix} \1 & 0 & 0 \\ 0 & \1 & 0 \\
0 & 0 & \e^{-\ii\sk\underline{a}}
\end{pmatrix},
\end{equation}
and
\begin{equation}\label{T:def:neu}
T(\sk;\underline{a}) := \begin{pmatrix} 0 & 0 & 0 \\ 0 & 0 &
\e^{\ii\sk\underline{a}} \\ 0 & \e^{\ii\sk\underline{a}} & 0
\end{pmatrix}
\end{equation}
with respect to the orthogonal decomposition \eqref{K:def}. The diagonal
$|\cI|\times |\cI|$ matrices $\e^{\pm \ii\sk\underline{a}}$ are given by
\begin{equation}
\label{diag} [\e^{\pm \ii\sk\underline{a}}]_{jk}=\delta_{jk}\e^{\pm \ii\sk
a_{j}}\quad
                       \text{for}\quad j,k\in\;\cI.
\end{equation}

\begin{lemma}\label{lem:Green}
For any maximal isotropic subspace $\cM\subset{}^d\cK$ the resolvent
\begin{equation*}
(-\Delta(\cM;\underline{a})-\sk^2)^{-1}\quad\text{for} \quad
\sk^2\in\C\setminus\mathrm{spec}(-\Delta(\cM;\underline{a}))\quad\text{with}\quad
\det(A+\ii\sk B)\neq 0
\end{equation*}
is the integral operator with the $(|\cI|+|\cE|)\times(|\cI|+|\cE|)$
matrix-valued integral kernel $r_{\cM}(x,y;\sk,\underline{a})$, $\Im\sk>0$,
admitting the representation
\begin{equation}\label{r:M:alternativ}
\begin{split}
& r_{\cM}(x,y;\sk,\underline{a})  = r^{(0)}(x,y,\sk)\\ & + \frac{\ii}{2\sk}
\Phi(x,\sk) R(\sk;\underline{a})^{-1}[\1-\mathfrak{S}(\sk;\cM)
T(\sk;\underline{a})]^{-1}\mathfrak{S}(\sk;\cM)R(\sk;\underline{a})^{-1}
\Phi(y,\sk)^T,
\end{split}
\end{equation}
where $R(\sk;\underline{a})$ is defined in \eqref{U:def:neu}, the matrix
$\Phi(x,\sk)$ is given by
\begin{equation*}
\Phi(x,\sk) := \begin{pmatrix}\phi(x,\sk) & 0 & 0 \\ 0 & \phi_+(x,\sk) &
\phi_-(x,\sk)
\end{pmatrix}
\end{equation*}
with diagonal matrices $\phi(x,\sk)=\diag\{\e^{\ii\sk x_j}\}_{j\in\cE}$,
$\phi_\pm(x,\sk)=\diag\{\e^{\pm\ii\sk x_j}\}_{j\in\cI}$, and
\begin{equation*}
[r^{(0)}(x,y,\sk)]_{j,j^\prime} = \ii\delta_{j,j^\prime}
\frac{\e^{\ii\sk|x_j-y_j|}}{2\sk},\quad x_j,y_j\in I_j.
\end{equation*}
If $\cI=\emptyset$, this representation simplifies to
\begin{equation*}
r_{\cM}(x,y,\sk) = r^{(0)}(x,y,\sk) + \frac{\ii}{2\sk} \phi(x,\sk)
\mathfrak{S}(\sk;\cM) \phi(y,\sk).
\end{equation*}
\end{lemma}

The integral kernel $r_{\cM}(x,y;\sk,\underline{a})$ is called
\emph{Green's function} or \emph{Green's matrix}.

The proof of Lemma \ref{lem:Green} is given in \cite{KS3}.

\subsection{Walks on Graphs and Cycles}

We recall the following definitions from \cite{KS3}. A nontrivial walk $\bw$
on the graph $\cG$ from $j\in\cE\cup\cI$ to $j^\prime\in\cE\cup\cI$ is an
ordered sequence formed out of edges and vertices
\begin{equation}\label{walk:def}
\{j,v_0,j_1,v_1,\ldots,j_n,v_n,j^\prime\}
\end{equation}
such that
\begin{itemize}
\item[(i)]{$j_1,\ldots,j_n\in\cI$;}
\item[(ii)]{the vertices $v_0\in V$ and $v_n\in V$ satisfy $v_0\in\partial(j)$,
$v_0\in\partial(j_1)$, $v_n\in\partial(j^\prime)$, and
$v_n\in\partial(j_n)$;}
\item[(iii)]{for any $k\in\{1,\ldots,n-1\}$ the vertex $v_k\in V$ satisfies $v_k\in\partial(j_k)$ and
$v_k\in\partial(j_{k+1})$;}
\item[(iv)]{$v_k=v_{k+1}$ for some $k\in\{0,\ldots,n-1\}$ if and only if $j_k$ is a tadpole.}
\end{itemize}
If $j,j^\prime\in\cE$ this definition is equivalent to that given in
\cite{KS4}.

The number $n$ is the \emph{combinatorial length} $|\bw|_{\mathrm{comb}}$
and the number
\begin{equation*}
|\bw|=\sum_{k=1}^n a_{j_k} >0
\end{equation*}
is the \emph{metric length} of the walk $\bw$.

A \emph{trivial} walk on the graph $\cG$ from $j\in\cE\cup\cI$ to
$j^\prime\in\cE\cup\cI$ is a triple $\{j,v,j^\prime\}$ such that
$v\in\partial(j)$ and $v\in\partial(j^\prime)$. Otherwise the walk is
called nontrivial. In particular, if $\partial(j)=\{v_0,v_1\}$, then
$\{j,v_0,j\}$ and $\{j,v_1,j\}$ are trivial walks, whereas
$\{j,v_0,j,v_1,j\}$ and $\{j,v_1,j,v_0,j\}$ are nontrivial walks of
combinatorial length $1$ and of metric length $a_j$. Both the combinatorial
and metric length of a trivial walk are zero.

We will say that the walk \eqref{walk:def} leaves the edge $j$ through the
vertex $v_0$ and enters the edge $j^\prime$ through the vertex $v_n$. A
trivial walk $\{j,v,j^\prime\}$ leaves $j$ and enters $j^\prime$ through
the same vertex $v$.

For any given walk $\bw$ from $j\in\cE\cup\cI$ to $j^\prime\in\cE\cup\cI$
we denote by $v_-(\bw)$ the vertex through which the walk leaves the edge
$j$ and by $v_+(\bw)$ the vertex through which the walk enters the edge
$j^\prime$. For trivial walks one has $v_-(\bw) = v_+(\bw)$.

Assume that the edges $j,j^\prime\in\cE\cup\cI$ are not tadpoles.
For a walk $\bw$ from $j$ to $j^\prime$ we set
\begin{equation*}
\dist(x_j, v_-(\bw)) := \begin{cases} x_j & \text{if}\quad
v_-(\bw)=\partial^-(j),\\ a_j - x_j & \text{if}\quad v_-(\bw)=\partial^+(j),
\end{cases}
\end{equation*}
and
\begin{equation*}
\dist(x_{j^\prime}, v_+(\bw)) := \begin{cases} x_{j^\prime} & \text{if}\quad
v_+(\bw)=\partial^-(j),\\ a_j - x_j & \text{if}\quad v_+(\bw)=\partial^+(j).
\end{cases}
\end{equation*}

A walk $\bw=\{j,v_0,j_1,v_1,\ldots,j_n,v_n,j^\prime\}$ \emph{traverses} an
internal edge $i\in\cI$ if $j_k=i$ for some $1\leq k \leq n$. It
\emph{visits} the vertex $v$ if $v_k=v$ for some $0\leq k \leq n$. The
\emph{score} $\underline{n}(\bw)$ of a walk $\bw$ is the set
$\{n_i(\bw)\}_{i\in\cI}$ with $n_i(\bw)\geq 0$ being the number of times
the walk $\bw$ traverses the internal edge $i\in\cI$. In particular,
\begin{equation*}
|\bw| = \sum_{i\in\cI} a_i n_i(\bw).
\end{equation*}

We say that the walk is \emph{transmitted} at the vertex $v_k$ if either
$v_k=\partial(e)$ or $v_k=\partial(e^\prime)$ or $v_k\in\partial(i_k)$,
$v_k\in\partial(i_{k+1})$, and $i_k\neq i_{k+1}$. We say that a trivial
walk from $e^\prime$ to $e$ is transmitted at the vertex
$v=\partial(e)=\partial(e^\prime)$ if $e\neq e^\prime$. Otherwise the walk
is said to be \emph{reflected}.

Let $\cW_{j,j^\prime}$, $j,j^\prime\in\cE\cup\cI$ be the (infinite if $\cI
\neq \emptyset$) set of all walks $\bw$ on $\cG$ from $j$ to $j^\prime$. By
$\cW_{j,j^\prime}(\underline{n})$, $\underline{n}\in(\N_0)^{|\cI|}$ we
denote set of all walks $\bw$ on $\cG$ from $j$ to $j^\prime$ with score
$\underline{n}$.

A walk
\begin{equation}\label{walk:ref}
\bw = \{j,v_0,j_1,v_1,\ldots,j_n,v_n,j^\prime\}
\end{equation}
is called \emph{closed} if $j=j^\prime$. It is called \emph{properly
closed} if it is closed and $v_-(\bw) \equiv v_0 \neq v_n \equiv v_+(\bw)$.
For any closed walk $\bw$ we denote by $j(\bw)$ its initial edge, that is,
$j(\bw)=j=j^\prime$.

For instance, let $j$ be an arbitrary internal edge with
$\partial(j)=\{v_0, v_1\}$, $v_0\neq v_1$. Then, the walk $\{j,v_0,j\}$ is
not properly closed, whereas $\{j,v_0,j,v_1,j\}$ is. Any closed walk from
an external edge is not properly closed.

We will say that two properly closed walks $\bw$ and $\bw^\prime$ are
equivalent, if they can be obtained from each other by successive
application of the transformation of the form
\begin{equation*}
\{j,v_0,j_1,v_1,\ldots,j_n,v_n,j\} \rightarrow
\{j_1,v_1,\ldots,j_n,v_n,j,v_0,j_1\}.
\end{equation*}

A \emph{cycle} is an equivalence class of properly closed walks.
We will say that the cycle $\fc$ is associated with a walk $\bw$ and write $\fc(\bw)$,
if $\bw$ is in the equivalence class $\fc$.

The number
\begin{equation}\label{def:length:cycle}
|\fc| := |\bw| + a_j,
\end{equation}
where $\bw$ is an arbitrary walk in the equivalence class $\fc$ and
$j=j(\bw)$, will be called the \emph{metric length} of the cycle $\fc$.
Obviously, this definition does not depend on the particular choice of the
walk $\bw$ in $\fc$. The set of all cycles on the graph $\cG$ will be
denoted by $\fC$.

We call a cycle $\fc$ \emph{primitive} if for any $\bw$ in $\fc$ there is no
integer $p\geq 2$ such that $\{p^{-1} n_i(\bw)\}_{i\in\cI}$ is a score of a
properly closed walk. For instance, if $j\in\cI$,
$\partial(j)=\{v_0,v_1\}$, $v_0\neq v_1$, the cycle associated with the
properly closed walk $\{j,v_0,j,v_1,j\}$ is primitive, whereas the cycle
associated with the properly closed walk $\{j,v_0,j,v_1,j,v_0,j,$ $v_1,
j\}$ is not.

For an arbitrary cycle $\fc$ and any $p\in\N$ we denote by $p\fc$ the unique
cycle with the following property: For any walk $\bw$ in $p\fc$ there is a
walk $\bw^\prime$ in $\fc$ with the score $\{p^{-1} n_i(\bw)\}_{i\in\cI}$.
The set of all primitive cycles on the graph $\cG$ will be denoted by
$\fC_{\mathrm{prim}}$.

The reverse of the walk $\bw$ is the walk $\bw_{\mathrm{rev}}$ is
$\{j^\prime,v_n, j_n,\ldots,v_1,j_1,v_0,j\}$. It may happen that
$\bw_{\mathrm{rev}}=\bw$. If $\bw$ is a properly closed walk, then its
reverse $\bw_{\mathrm{rev}}$ is also properly closed. We will write
$\fc_{\mathrm{rev}}$ for the equivalence class associated with
$\bw_{\mathrm{rev}}$ for $\bw$ in $\fc$. Obviously, the map
$\fc\mapsto\fc_{\mathrm{rev}}$ satisfies $(p\fc)_{\mathrm{rev}}=p
\fc_{\mathrm{rev}}$ for any $p\in\N$. From what has been just said, it
follows that the case $\fc = \fc_{\mathrm{rev}}$ may occur.

\subsection{Combinatorial Expansion of the Resolvent}

For any $\sk\in\C$ with $\Im \sk >0$, the operator $T(\sk;\underline{a})$ defined in \eqref{T:def:neu}
is a uniform contraction. Therefore,
\begin{equation}\label{series}
[\1-\mathfrak{S}(\sk;\cM)
T(\sk;\underline{a})]^{-1}\mathfrak{S}(\sk;\cM)=\sum_{n=0}^\infty
\mathfrak{S}(\sk;\cM)\left(T(\sk;\underline{a})\mathfrak{S}(\sk;\cM)\right)^n
\end{equation}
converges uniformly in $\sk$ for
$\sk$ in any of the sets $\{\sk\in\C|\Im\sk >\epsilon>0\}$. Inserting
\eqref{series} into \eqref{r:M:alternativ}, we get
\begin{equation*}
\begin{split}
& r_{\cM}(x,y;\sk,\underline{a}) = r^{(0)}(x,y,\sk)\\
& +\frac{\ii}{2\sk} \sum_{n=0}^\infty \Phi(x,\sk) R(\sk;\underline{a})^{-1}
\mathfrak{S}(\sk;\cM)\left(T(\sk;\underline{a})\mathfrak{S}(\sk;\cM)\right)^n
R(\sk;\underline{a})^{-1}\Phi(y,\sk)^T.
\end{split}
\end{equation*}
For maximal isotropic subspaces satisfying any of the equivalent conditions
of Proposition \ref{k-independent}, $\mathfrak{S}(\sk;\cM)$ is independent
of $\sk$. Thus, using \eqref{series} we get

\begin{proposition}\label{propo:res:comb}
Assume that the graph $\cG$ has no tadpoles. For any
maximal iso\-tropic subspace $\cM$ satisfying any of the equivalent
conditions of Proposition \ref{k-independent}, the Green function of the
Laplace operator $\Delta(\cM,\underline{a})$ has the absolutely converging
expansion
\begin{equation}\label{res:comb}
\begin{split}
& [r_{\cM}(x,y;\sk,\underline{a})]_{j,j^\prime} = \frac{\ii}{2\sk}
\delta_{j,j^\prime} \e^{\ii\sk|x_j-y_j|}\\
& +\frac{\ii}{2\sk} \sum_{\bw\in\cW_{j,j^\prime}} \e^{\ii\sk \dist(x_j,
v_-(\bw))} W_{\cM}(\bw) \e^{\ii\sk|\bw|} \e^{\ii\sk \dist(y_{j^\prime},
v_+(\bw))},\qquad \Im\sk>0,
\end{split}
\end{equation}
where $W_{\cM}(\bw)$ is a (complex-valued) weight of the walk
$\bw=\{j,v_0,j_1,v_1,\ldots,j_n,v_n,j^\prime\}$,
\begin{equation}\label{weight:2}
W_{\cM}(\bw) = \prod_{l=0}^{|\bw|_{\mathrm{comb}}} [\fS(A(v_l),
B(v_l))]_{i_{l+1}, i_{l}}.
\end{equation}
\end{proposition}

\subsection{Heat Kernel}

The semigroup generated by the positive operator $-\Delta(\cM,
\underline{a})$ is related to its resolvent by the Dunford-Taylor integral
(see \cite[Section IX.1.6]{Kato})
\begin{equation*}
\e^{t\Delta(\cM, \underline{a})} = -\frac{1}{2\pi\ii} \int_\gamma
\e^{-t\lambda} (-\Delta(\cM, \underline{a})-\lambda)^{-1} d\lambda,
\end{equation*}
where $\gamma$ is any contour encircling a positive semiline
counterclockwise. The integral converges in the sense of Bochner.
Using the well-known identity
\begin{equation*}
\frac{1}{2\pi}\int_{-\infty+\ii\epsilon}^{+\infty+\ii\epsilon} \e^{-\sk^2
t} \e^{\ii\sk u} d\sk = g_t(u):=\frac{1}{\sqrt{4\pi t}}
\exp\left\{-u^2/4t\right\},\qquad \epsilon>0,
\end{equation*}
we immediately get the following corollary of Proposition \ref{propo:res:comb}.

\begin{corollary}\label{6:cor:1}
Assume that the maximal isotropic subspace $\cM$ satisfies any of the
equivalent conditions of Proposition \ref{k-independent} and defines local
boundary conditions on the graph $\cG$. Assume, in addition, that the graph
$\cG$ has no tadpoles. Then the heat kernel of
$-\Delta(\cM,\underline{a})$ has the absolutely convergent expansion
\begin{equation}\label{heatseries}
\begin{split}
& [p_t(x,y;\cM,\underline{a})]_{j,j^\prime} = \delta_{j,j^\prime}
g_t(x_j-y_{j})\\ &\quad + \sum_{\bw \in \cW_{j,j^\prime}} W_{\cM}(\bw)\,
g_t\big(\dist(x_j, v_-(\bw))+|\bw|+\dist(y_{j^\prime}, v_+(\bw))\big),
\end{split}
\end{equation}
The series converges uniformly in $\displaystyle x,y\in\Bigtimes_{j\in\cE\cup\cI} I_j$.
\end{corollary}

\begin{proof}
It remains to prove that the series in \eqref{heatseries} converges uniformly.
This follows from the estimate
\begin{equation*}
\begin{aligned}
& \left|\sum_{\bw \in \cW_{j,j^\prime}} W_{\cM}(\bw)\, g_t\big(\dist(x_j,
v_-(\bw))+|\bw|+\dist(x_{j^\prime}, v_+(\bw))\big)\right|\\
&\qquad\leq \frac{1}{\sqrt{4\pi t}}
\sum_{\bw \in \cW_{j,j^\prime}} \exp\{-|\bw|^2/4t\}\\
&\qquad\leq\frac{1}{\sqrt{4\pi t}}
\sum_{\underline{n}\in(\N_0)^{|\cI|}}\sum_{\bw \in
\cW_{j,j^\prime}(\underline{n})}
\exp\{-|\bw|^2/4t\}\\
&\qquad\leq \frac{1}{\sqrt{4\pi t}}\sum_{\underline{n}\in(\N_0)^{|\cI|}}
\frac{|\underline{n}|!}{\prod_{i\in\cI}n_i!}
\exp\{-|\underline{n}|^2 a_{\min}^2/4t\}\\
&\qquad\leq \frac{1}{\sqrt{4\pi t}}\sum_{n=0}^\infty |\cI|^n
\exp\{-n^2\,a_{\min}^2/4t\} < \infty,
\end{aligned}
\end{equation*}
where $\displaystyle a_{\min}=\min_{i\in\cI}\{a_j\}$.
\end{proof}

In the particular case of a connected graph with $\cI=\emptyset$ and
standard boundary conditions, we observe that $\cW_{j,j^\prime}$ consists of
precisely one walk. Hence, from \eqref{heatseries} we get the representation
(7.1) in \cite{KS3}, which has first been derived in \cite{Gaveau:2} by
different methods. In the particular case $\cM=\cM_{\mathrm{st}}$ for
compact graphs $\cG$ a representation similar to \eqref{heatseries} has been
obtained by Roth in \cite{Roth:2}.

We will now look at the situation with standard boundary conditions at all
vertices in more detail. For a given walk $\bw$ we set
\begin{equation*}
\begin{aligned}
N_{\mathrm{refl}}(\bw)&=\text{number of times the walk $\bw$ is reflected},\\
N_{\mathrm{trans}}(\bw)&=\text{number of times the walk $\bw$ is
transmitted},
\end{aligned}
\end{equation*}
such that
\begin{equation*}
N_{\mathrm{refl}}(\bw)+N_{\mathrm{trans}}(\bw)= |\bw|_{\mathrm{comb}}+1.
\end{equation*}
{}From Corollary \ref{6:cor:1} we obtain

\begin{corollary}\label{6:cor:2}
Assume that the graph $\cG$ is $k$-regular, that is, $\deg(v)=k$ for all
$v\in V$, and has no tadpoles.  Then for standard
boundary conditions at each of the vertices the heat kernel of
$-\Delta(\cM_{\mathrm{st}},\underline{a})$ has the absolutely convergent
expansion
\begin{equation}\label{reftrans2}
\begin{split}
[p_t(x,y;\cM_{\mathrm{st}},\underline{a})]_{j,j^\prime} &
=\delta_{j,j^\prime} g_t(x_j-y_j)\\  & + \sum_{\bw \in \cW_{j,j^\prime}}
\left(\frac{2-k}{k}\right)^{N_{\mathrm{refl}}(\bw)}
\left(\frac{2}{k}\right)^{N_{\mathrm{trans}}(\bw)} \\ & \cdot g_t(\dist(x_j,
v_-(\bw))+|\bw|+\dist(y_{j^\prime}, v_+(\bw))).
\end{split}
\end{equation}
\end{corollary}

\section{The Trace Formula}\label{sec:trac:form}

On the exterior $\cG_{\mathrm{ext}}=(\partial
V,\emptyset,\cE,\partial|_{\cE})$ of the graph $\cG=(V,\cI,\cE,\partial)$ we
consider the Laplace operators $\Delta_+:=\Delta(A_{\cE}=0, B_{\cE}=\1)$
corresponding to Neumann boundary conditions and
$\Delta_-:=\Delta(A_{\cE}=\1, B_{\cE}=0)$ corresponding to Dirichlet
boundary conditions.

Let $\cJ:\; \cH_{\cE}\rightarrow \cH$ be the embedding operator defined for
any $\chi\in\cH_{\cE}$ by $\cJ\chi=\chi\oplus 0$, where the orthogonal sum
is taken with respect to the decomposition $\cH=\cH_{\cE}\oplus\cH_{\cI}$, such that $\cJ^\dagger \cJ$ is the identity on $\cH_{\cE}$ and $\cJ\cJ^\dagger$ the orthogonal projection in $\cH$ onto $\cH_{\cE}$.
If $\cE=\emptyset$, we set $\cJ=0$.

\begin{theorem}\label{thm:4:1}
Assume that the graph $\cG$ has no tadpoles. Let the maximal isotropic
subspace $\cM$ satisfy any of the equivalent conditions of Proposition
\ref{k-independent} and assume that it defines local boundary conditions on
$\cG$. Then
\begin{equation*}
\begin{split}
\tr_{\cH}\left(\e^{t\Delta(\cM,\underline{a})} - \cJ \e^{t\Delta_\pm}
\cJ^\dagger \right) & = \frac{L}{2\sqrt{\pi t}} + \frac{1}{4} \tr_{\cK}
\fS(\cM) \mp \frac{|\cE|}{4} \\ & + \frac{1}{2\sqrt{\pi t}}
\sum_{\fc\in\fC_{\mathrm{prim}}} \sum_{p\in\N} W_{\cM}(\fc)^p
|\fc|\exp\left\{-\frac{p^2 |\fc|^2}{4t}\right\},\quad t>0,
\end{split}
\end{equation*}
where $L:=\sum_{j\in\cI} a_j$ is the total metric length of the interior of
the graph $\cG$, and $W_{\cM}(\fc)$ is the weight $W_{\cM}(\bw)$ associated
with any walk in the cycle $\fc$. In particular, if the maximal isotropic
subspace $\cM$ corresponds to a magnetic perturbation of the standard
boundary conditions (see Example \ref{3:ex:mag}), then
\begin{equation}\label{gutzw}
\begin{split}
& \tr_{\cH}\left(\e^{t\Delta(\cM,\underline{a})} - \cJ \e^{t\Delta_\pm} \cJ^\dagger
\right) = \frac{L}{2\sqrt{\pi t}} + \begin{cases}\frac{|V|-|\cI|-|\cE|}{2} & \text{for ``$+$''} \\
\frac{|V|-|\cI|}{2} & \text{for ``$-$''} \end{cases} \\ & +
\frac{1}{2\sqrt{\pi t}} \sum_{\fc\in\fC_{\mathrm{prim}}} \sum_{p\in\N}
W_{\cM_{\mathrm{st}}}(\fc)^p |\fc|\exp\left\{-\frac{p^2 |\fc|^2}{4t}\right\}
\e^{\ii p\Phi(\fc)},\quad t>0,
\end{split}
\end{equation}
where $\Phi(\fc)$ is the magnetic flux through the cycle $\fc$ defined in
\eqref{def:flux} below.
\end{theorem}

\begin{remark}
Since $\Phi(\fc_{\mathrm{rev}})=-\Phi(\fc)$ and
$W_{\cM_{\mathrm{st}}}(\fc_{\mathrm{rev}})=W_{\cM_{\mathrm{st}}}(\fc)$ (see \eqref{flux:rev} below),
the factor $\e^{\ii p\Phi(\fc)}$ in \eqref{gutzw} can be
replaced by $\cos(p\Phi(\fc))$.
\end{remark}

Before we turn to the proof of Theorem \ref{thm:4:1} we will briefly
discuss the trace formula \eqref{gutzw}. The first term on its r.h.s.\ is a
familiar Weyl term. In complete analogy with small $t$ expansion of the
trace of heat semigroups on smooth two-dimensional Riemannian manifolds
\cite{McKean:Singer}, the second term depends solely on the topology of the
graph: The number $|\cI|-|V|$ is the Euler characteristic of the graph
viewed as a simplicial complex. In the context of metric graphs the Euler
characteristics has been discussed in \cite{KS5} and \cite{KuKu}. If we
interpret the quantity $\frac{1}{2}\deg(v)-1$ as the local curvature at the
vertex $v\in V$, then \eqref{1:thm:gt} gives a discrete version of the
Gau{\ss}-Bonnet theorem for compact graphs:
\begin{equation*}
\sum_{v\in V} \left(\frac{1}{2}\deg(v)-1\right) = |\cI|-|V|.
\end{equation*}
We emphasize that the local curvature at the vertices of the graph is not a
curvature in the sense of Regge calculus \cite{Regge}. Regge calculus,
however, can be used to define local curvatures on piecewise flat (or
piecewise linear) spaces including Lipschitz-Killing curvatures and
boundary curvatures \cite{CMS}. In particular, these curvatures have been
used in \cite{CMS} to give an alternative proof of the Chern-Gau{\ss}-Bonnet
theorem for compact closed Riemannian manifolds.

In a similar vein the term $|\cI|+|\cE|-|V|$ appearing in \eqref{gutzw} can
be interpreted as the relative Euler characteristic (cf.~\cite{RS}) of the
graph $\cG$ whenever $\cG$ is noncompact, that is, when $\cE\neq\emptyset$.
In the context of exterior domains in $\R^d$, the relation of Laplace
operators on forms with absolute and relative boundary conditions (analogs
of Dirichlet and Neumann boundary conditions) to absolute and relative
Euler characteristics, respectively, has been established in \cite{BMS} as
a relative index theorem in the spirit of \cite{GL}.

The sum over primitive cycles of the graph $\cG$ in the r.h.s.\ of
\eqref{gutzw} is an analog of the sum over primitive periodic geodesics on
the manifold in the celebrated Selberg trace formula \cite{Selberg} (see
also \cite{Comtet}, \cite{Gutzwiller}, \cite{Hejhal}, \cite{Levay},
\cite{McKean}).

The remainder of this section is devoted to the proof of Theorem
\ref{thm:4:1}.

The heat semigroups $\e^{t\Delta_\pm}$ are integral operators with kernels
\begin{equation*}
h^{\pm}_t(x_j,y_j) := g_t(x_j-y_j) \pm g_t(x_j+y_j),
\end{equation*}
respectively. That the difference $\e^{t\Delta(\cM,\underline{a})} - \cJ
\e^{t\Delta_\pm} \cJ^\dagger$ is trace class follows from the fact that
$\Delta(\cM,\underline{a})$ is a finite rank perturbation of $\Delta_\pm$.
For any trace class operator $K$ on $\cH$
\begin{equation*}
\tr_{\cH} K = \sum_{j\in\cI\cup\cE} \tr_{\cH_j} P_j K P_j,
\end{equation*}
where $P_j$ is the orthogonal projection in $\cH$ onto $\cH_j$. Observe that
$P_j\left(\e^{t\Delta(\cM,\underline{a})}-\e^{t\Delta_\pm}\right)P_j$ are
integral operators on $L^2(I_j)$ with kernels jointly continuous in
$x_j,y_j\in \overset{o}{I}_j$ (due to the uniform convergence of the series
in \eqref{heatseries}). Therefore, by Corollary III.10.2 in
\cite{Gohberg:Krein}, the trace of
$P_j\left(\e^{t\Delta(\cM,\underline{a})}-\e^{t\Delta_\pm}\right)P_j$
equals the integral of its kernel over the diagonal. Hence, from Corollary
\ref{6:cor:1}, we get
\begin{equation}\label{3:9:neu}
\begin{split}
& \tr_{\cH}\left(\e^{t\Delta(\cM)} - \cJ \e^{t\Delta_\pm} \cJ^\dagger\right)\\
&\quad = \sum_{j\in\cI} \int_{I_j} g_t(0) dx_j + \sum_{j\in\cE} \int_{I_j}
[g_t(0)-h_t^\pm(x_j,x_j)] dx_j\\ &\quad + \sum_{j\in\cI\cup\cE}
\sum_{\bw\in\cW_{j,j}} W(\bw) \int_{I_j} g_t(\dist(x_j,
v_-(\bw))+|\bw|+\dist(x_j, v_+(\bw))) dx_j,
\end{split}
\end{equation}
where the sum converges absolutely. We will evaluate the different
contributions to the r.h.s.\ of \eqref{3:9:neu} separately. Essentially we
will follow the original ideas of Roth developed in \cite{Roth:2}.

\textbf{1.} We start with the terms in \eqref{3:9:neu} not associated with
any walk on the graph $\cG$. Simple calculations yield
\begin{equation*}
\int_0^{a_j} g_t(0) dx_j = \frac{a_j}{\sqrt{4\pi t}}\qquad\text{if}\qquad
j\in\cI
\end{equation*}
and
\begin{equation*}
\int_0^{\infty} \left[g_t(0)-h_t^\pm(x_j,x_j)\right] dx_j = \mp
\int_0^{\infty} g_t(2x_j) dx_j = \mp\frac{1}{4} \qquad\text{if}\qquad
j\in\cE.
\end{equation*}
Summing over all edges $j\in\cI\cup\cE$ we get the following contribution
to \eqref{3:9:neu}
\begin{equation*}
\frac{1}{\sqrt{4\pi t}} \sum_{j\in\cI} a_j \mp \frac{|\cE|}{4}.
\end{equation*}

\textbf{2.} Next we study the contributions from properly closed walks. Let
$\bw$ be a (nontrivial) properly closed walk from $j\in\cI$ to $j\in\cI$. In
this case $v_-(\bw) \neq v_+(\bw)$ and, therefore, we have
\begin{equation*}
\dist(x_j, v_-(\bw)) + \dist(x_j, v_+(\bw)) = a_j.
\end{equation*}
Therefore,
\begin{equation*}
\sum_{\text{properly closed}\, \bw\in\cW_{j,j}} W_{\cM}(\bw) \int_{I_j}
g_t(|\bw|+a_j) dx_j = a_j W_{\cM}(\bw) g_t(|\bw|+a_j)
\end{equation*}
Summing over all walks in the cycle $\fc=\fc(\bw)$ and using
\eqref{def:length:cycle} we get
\begin{equation*}
W_{\cM}(\fc) g_t(|\fc|) \sum_{\bw\in\fc} a_{j(\bw)}.
\end{equation*}
Obviously, $\displaystyle\sum_{\bw\in\fc} a_{j(\bw)}=|\fc^\prime|$ if
$\fc=p\fc^\prime$ for some $p\in\N$ and some primitive cycle $\fc^\prime$.
Thus, the sum of the contributions in \eqref{3:9:neu} from all properly
closed walks equals
\begin{equation*}
\frac{1}{2\sqrt{\pi t}} \sum_{\fc\in\fC_{\mathrm{prim}}} \sum_{p\in\N}
W_{\cM}(p\fc) |\fc|\exp\left\{-\frac{p^2 |\fc|^2}{4t}\right\},\quad t>0.
\end{equation*}
Obviously, the relation $W_{\cM}(p\fc)=W_{\cM}(\fc)^p$ holds for all
$p\in\N$.

\textbf{3.} We turn to the contributions which are not coming from properly
closed walks. In this case $v_-(\bw) = v_+(\bw)$ and, therefore, we have
\begin{equation*}
\dist(x_j, v_-(\bw)) + \dist(x_j, v_+(\bw)) = \begin{cases} 2x_j &
\text{if}\quad v_-(\bw)\in\partial^-(j),\\ 2(a_j-x_j) & \text{if}\quad v_-(\bw)\in\partial^+(j).
\end{cases}
\end{equation*}

We will call a not properly closed walk $\bw$ a walk of type $A$ if it is of
the form
\begin{equation*}
\bw=\{j_p, v_p, j_{p-1},v_{p-1},\ldots, j_0,v_0,j_0,\ldots,v_{p-1}, j_{p-1}, v_p, j_p\}.
\end{equation*}
Otherwise a not properly closed walk $\bw$ is called a walk of type $B$. By
$\cW_{j,j}^A$ and $\cW_{j,j}^B$ we denote the set of all walks from $j$ to
$j$ of type $A$ and $B$, respectively.

Obviously, any not properly closed walk is either of type $A$ or $B$. Any
walk of type $A$ is invariant with respect to reversion, that is,
$\bw_{\mathrm{rev}}=\bw$, whereas walks of type $B$ are not.

The following two lemmas complete the proof of the first part of Theorem
\ref{thm:4:1}.

\begin{lemma}\label{lem:4:6}
\begin{equation*}
\sum_{j\in\cI\cup\cE} \sum_{\bw\in \cW_{j,j}^A} W_{\cM}(\bw) \int_{I_j}
g_t(2x_j + |\bw|) dx_j  = \frac{1}{4} \tr_{\cK}\fS(\cM).
\end{equation*}
\end{lemma}

\begin{lemma}\label{lem:4:5}
\begin{equation*}
\sum_{j\in\cI\cup\cE} \sum_{\bw\in \cW_{j,j}^B} W_{\cM}(\bw) \int_{I_j}
g_t(2x_j + |\bw|) dx_j = 0.
\end{equation*}
\end{lemma}

\begin{proof}[Proof of Lemma \ref{lem:4:6}]
For an arbitrary vertex $v_0\in V$ and arbitrary $p\in\N_0$ we set
\begin{equation*}
\begin{split}
\partial G_{v_0}^{\cI}(p) & := \{\text{walks of the form
$\{j_p,v_p,j_{p-1},\ldots,j_0,v_0,j_0,\ldots,j_{p-1},v_p,j_p\}$}\\ &
\qquad\qquad\qquad\qquad\qquad\qquad\qquad\qquad\qquad\qquad\qquad\qquad
\text{with $j_p\in\cI$}\},\\
\partial G_{v_0}^{\cE}(p) & := \{\text{walks of the form
$\{j_p,v_p,j_{p-1},\ldots,j_0,v_0,j_0,\ldots,j_{p-1},v_p,j_p\}$}\\ &
\qquad\qquad\qquad\qquad\qquad\qquad\qquad\qquad\qquad\qquad\qquad\qquad
\text{with $j_p\in\cE$}\},\\
\partial G_{v_0}(p) & := \partial G_{v_0}^{\cI}(p) \cup \partial
G_{v_0}^{\cE}(p),\qquad \text{and}\qquad G_{v_0}(p) := \bigcup_{q=0}^p \partial
G_{v_0}(q).
\end{split}
\end{equation*}
We claim that
\begin{equation}\label{xxx}
\begin{split}
& \sum_{\bw\in G_{v_0}(p)} W_{\cM}(\bw) \int_{I_{j(\bw)}} g_t(2 x_{j(\bw)}+|\bw|)
dx_{j(\bw)} =\frac{1}{4}\tr_{\cL_{v_0}}[\fS_{v_0}]\\ & \qquad -\frac{1}{4}
\sum_{\bw\in\partial G_{v_0}^{\cI}(p)} W_{\cM}(\bw) \erfc\left(\frac{|\bw|+2
a_{j(\bw)}}{2\sqrt{t}}\right)
\end{split}
\end{equation}
holds for all $p\in\N_0$. Here for brevity we set
$\fS_{v_{0}}:=\fS(\cM_{v_{0}})$, where $\cM_{v_{0}}\subset{}^d\cL_{v_0}$ is
the maximal isotropic subspace from the orthogonal decomposition
\eqref{propo:ortho}.
\begin{equation*}
\erfc(s):=\frac{2}{\sqrt{\pi}}\int_s^\infty \e^{-u^2}\;du
\end{equation*}
denotes the complementary error function \cite{AS}. The proof is by
induction. For $p=0$ we have
\begin{equation*}
\begin{split}
\sum_{\bw\in G_{v_0}(0)} W_{\cM}(\bw) \int_{I_{j(\bw)}} g_t(2
x_{j(\bw)}+|\bw|) dx_{j(\bw)} & =
\frac{1}{4}\tr_{\cL_{v_0}}[\fS_{v_0}]\\
-\frac{1}{4} \sum_{\bw\in\partial G_{v_0}^{\cI}(0)} W_{\cM}(\bw)
\erfc\left(\frac{a_{j(\bw)}}{\sqrt{t}}\right).
\end{split}
\end{equation*}
Now assume that \eqref{xxx} holds for some $p\in\N$ and consider
\begin{equation*}
\begin{split}
& \sum_{\bw\in G_{v_0}(p+1)} W_{\cM}(\bw) \int_{I_{j(\bw)}} g_t(2 x_{j(\bw)}+|\bw|) dx_{j(\bw)} \\
&= \sum_{\bw\in G_{v_0}(p)} W_{\cM}(\bw) \int_{I_{j(\bw)}} g_t(2 x_{j(\bw)}+|\bw|) dx_{j(\bw)} \\
& + \sum_{\bw\in \partial G_{v_0}(p+1)} W_{\cM}(\bw) \int_{I_{j(\bw)}} g_t(2 x_{j(\bw)}+|\bw|) dx_{j(\bw)}\\
& = \frac{1}{4}\tr_{\cL_{v_0}}[\fS_{v_0}] -\frac{1}{4} \sum_{\bw\in\partial G_{v_0}^{\cI}(p)} W_{\cM}(\bw) \erfc\left(\frac{|\bw|+2 a_{j(\bw)}}{2\sqrt{t}}\right)\\
& + \frac{1}{4} \sum_{\bw\in \partial G_{v_0}^{\cE}(p+1)} W_{\cM}(\bw) \erfc\left(\frac{|\bw|}{2\sqrt{t}}\right)\\
& +\frac{1}{4} \sum_{\bw\in\partial G_{v_0}^{\cI}(p+1)} W_{\cM}(\bw)
\left[\erfc\left(\frac{|\bw|}{2\sqrt{t}}\right)-\erfc\left(\frac{|\bw|+2 a_{j(\bw)}}{2\sqrt{t}}\right)\right]\\
&= \frac{1}{4}\tr_{\cL_{v_0}}[\fS_{v_0}] -\frac{1}{4} \sum_{\bw\in\partial G_{v_0}^{\cI}(p)} W_{\cM}(\bw) \erfc\left(\frac{|\bw| +2 a_{j(\bw)}}{2\sqrt{t}}\right)\\
& +\frac{1}{4} \sum_{\bw\in\partial G_{v_0}(p+1)} W_{\cM}(\bw)
\erfc\left(\frac{|\bw|}{2\sqrt{t}}\right)\\ & -\frac{1}{4} \sum_{\bw\in\partial
G_{v_0}^{\cI}(p+1)} W_{\cM}(\bw) \erfc\left(\frac{|\bw|+2 a_{j(\bw)}}{2\sqrt{t}}\right).
\end{split}
\end{equation*}
It remains to prove that the sum of the second and third terms on the r.h.s.\ is zero. Let
$\bw\in\partial G_{v_0}(p+1)$ be arbitrary. Write the walk $\bw$ as $\bw = \{j_{p+1}, v_{p+1},
j_p,\ldots, j_0, v_0, j_0, \ldots, j_p,$ $v_{p+1}, j_{p+1}\}$. Then
$\bw^\prime :=\{j_p,\ldots, j_0, v_0, j_0, \ldots, j_p\}\in\partial
G_{v_0}(p)$ with $j_p\in\cI$. Hence,
\begin{equation*}
W_{\cM}(\bw)=[\fS_{v_{p+1}}]_{j_p,j_{p+1}} [\fS_{v_{p+1}}]_{j_{p+1},j_p}
W_{\cM}(\bw^\prime)
\end{equation*}
and $|\bw|=|\bw^\prime|+2a_{j(\bw^\prime)}$. Thus,
\begin{equation*}
\begin{split}
& \sum_{\bw\in\partial G_{v_0}(p+1)} W_{\cM}(\bw)
\erfc\left(\frac{|\bw|}{2\sqrt{t}}\right) \\ & \quad =
\sum_{\bw^\prime\in\partial G_{v_0}^{\cI}(p)} W_{\cM}(\bw^\prime)
\erfc\left(\frac{|\bw^\prime|+2a_{j(\bw^\prime)}}{2\sqrt{t}}\right)\\
& \quad\quad\cdot\sum_{j_{p+1}\in\cS(v_{p+1})} [\fS_{v_{p+1}}]_{j_p,j_{p+1}}
[\fS_{v_{p+1}}]_{j_{p+1},j_p}.
\end{split}
\end{equation*}
By (iii) in Proposition \ref{k-independent}, we have $\fS_{v_{p+1}}^2=\1$
and, therefore,
\begin{equation*}
\sum_{j_{p+1}\in\cS(v_{p+1})} [\fS_{v_{p+1}}]_{j_p,j_{p+1}}
[\fS_{v_{p+1}}]_{j_{p+1},j_p} = [\fS_{v_{p+1}}^2]_{j_p,j_{p}} = 1,
\end{equation*}
which completes the proof of \eqref{xxx}.

By the absolute convergence of the series \eqref{3:9:neu}, from \eqref{xxx}
it follows that
\begin{equation*}
\lim_{p\to\infty} \sum_{\bw\in G_{v_0}(p)} W_{\cM}(\bw) \int_{I_{j(\bw)}}
g_t(2x_{j(\bw)} + |\bw|) dx_{j(\bw)} = \frac{1}{4} \tr_{\cL_{v_0}}
\fS_{v_0}.
\end{equation*}
Observing that
\begin{equation*}
\begin{split}
& \sum_{j\in\cI\cup\cE} \sum_{\bw\in\cW_{j,j}^A} W_\cM(\bw) \int_{I_j} g_t(2x_j +|\bw|) dx_j\\
& =\lim_{p\to\infty}
\sum_{v_0\in V}\sum_{\bw\in G_{v_0}(p)} W_{\cM}(\bw) \int_{I_{j(\bw)}} g_t(2 x_{j(\bw)}+|\bw|)
dx_{j(\bw)},
\end{split}
\end{equation*}
we obtain the claim of the lemma.
\end{proof}

\begin{proof}[Proof of Lemma \ref{lem:4:5}]
Any not properly closed walk of type $B$ is obviously of the form
\begin{equation*}
\{j_p, v_p, j_{p-1},\ldots,j_0,v_0, \fs, v_0,j_0, \ldots, j_{p-1}, v_p,
j_p\}
\end{equation*}
for some $p\in\N_0$, where $\fs$ stands for the sequence of internal edges and vertices
$i_1^\prime, v_1^\prime,\ldots,$ $v_n^\prime,i_n^\prime$ with $i_1^\prime\neq
i_n^\prime$. For an arbitrary $p\in\N_0$ we set
\begin{equation*}
\begin{split}
\partial F_{\fs, v_0}^{\cI}(p) & :=\{\text{walks of the form}\,\{j_p, v_p, j_{p-1},\ldots,j_0,v_0, \fs, v_0,j_0, \ldots, j_{p-1}, v_p,
j_p\}\\  &
\qquad\qquad\qquad\qquad\qquad\qquad\qquad\qquad\qquad\qquad\qquad\qquad\text{with}\quad
j_p\in\cI\},\\
\partial F_{\fs, v_0}^{\cE}(p) & :=\{\text{walks of the form}\,\{j_p, v_p, j_{p-1},\ldots,j_0,v_0, \fs, v_0,j_0, \ldots, j_{p-1}, v_p,
j_p\}\\  &
\qquad\qquad\qquad\qquad\qquad\qquad\qquad\qquad\qquad\qquad\qquad\qquad\text{with}\quad
j_p\in\cE\},\\
\partial F_{\fs, v_0}(p) & :=\partial F_{\fs, v_0}^{\cI}(p)\cup \partial
F_{\fs, v_0}^{\cE}(p),\qquad \text{and} \qquad \displaystyle F_{\fs,
v_0}(p):=\bigcup_{q=0}^p
\partial F_{\fs, v_0}(q).
\end{split}
\end{equation*}
We claim that
\begin{equation}\label{xxx:2}
\begin{split}
& \sum_{\bw\in F_{\fs, v_0}(p)} W_{\cM}(\bw) \int_{I_j(\bw)}
g_t(2x_{j(\bw)}+|\bw|) dx_{j(\bw)}\\ & \qquad = - \frac{1}{4}
\sum_{\bw\in\partial F_{\fs, v_0}^{\cI}(p)} W_{\cM}(\bw)
\erfc\left(\frac{|\bw|+2a_{j(\bw)}}{2\sqrt{t}}\right).
\end{split}
\end{equation}
The proof is again by induction. For $p=0$ we have
\begin{equation}\label{claim:p=0}
\begin{split}
& \sum_{\bw\in F_{\fs, v_0}(0)} W_{\cM}(\bw) \int_{I_{j(\bw)}}
g_t(2x_{j(\bw)}+|\bw|) dx_{j(\bw)}\\ & = \sum_{\bw\in \partial F_{\fs,
v_0}^{\cE}(0)} W_{\cM}(\bw) \int_{I_{j(\bw)}} g_t(2x_{j(\bw)} + |\bw|) dx_{j(\bw)}\\
& + \sum_{\bw\in \partial F_{\fs, v_0}^{\cI}(0)} W_{\cM}(\bw)
\int_{I_{j(\bw)}} g_t(2x_{j(\bw)} + |\bw|) dx_{j(\bw)}.
\end{split}
\end{equation}
Obviously, for any $\bw\in F_{\fs, v_0}(0)$ we have
$W_{\cM}(\bw)=[\fS_{v_0}]_{j,i_1^\prime} [\fS_{v_0}]_{i_n^\prime,j}
W^\prime_{\cM}$, where $W^\prime_{\cM}$ is a weight associated with the
sequence $\fs$ and $j=j(\bw)$. Therefore, if $j\in\cE$, then
\begin{equation*}
W_{\cM}(\bw) \int_{I_j} g_t(2x_{j} + |\bw|) dx_{j} = \frac{1}{4}
[\fS_{v_0}]_{j,i_1^\prime} [\fS_{v_0}]_{i_n^\prime,j} W^\prime_{\cM}
\erfc\left(\frac{|\bw|}{2\sqrt{t}}\right),
\end{equation*}
and, if $j\in\cI$, then
\begin{equation*}
\begin{split}
& W_{\cM}(\bw) \int_{I_j} g_t(2x_{j} + |\bw|) dx_{j} \\ & = \frac{1}{4}
[\fS_{v_0}]_{j,i_1^\prime} [\fS_{v_0}]_{i_n^\prime,j} W^\prime_{\cM}
\left(\erfc\left(\frac{|\bw|}{2\sqrt{t}}\right)-\erfc\left(\frac{|\bw|+2a_j}{2\sqrt{t}}\right)\right).
\end{split}
\end{equation*}
Again we use $\fS_{v_0}^2 = \1$, which in combination with $i_1^\prime\neq i_n^\prime$ gives
\begin{equation*}
\begin{split}
& \frac{1}{4} W^\prime_{\cM} \erfc\left(\frac{|\bw|}{2\sqrt{t}}\right)
\sum_{j\in\cS(v_0)} [\fS_{v_0}]_{j,i^\prime_1} [\fS_{v_0}]_{i^\prime_n,j} \\
& =\frac{1}{4} W^\prime_{\cM} \erfc\left(\frac{|\bw|}{2\sqrt{t}}\right)
\sum_{j\in\cS(v_0)} [\fS_{v_0}^2]_{i^\prime_n,i^\prime_1} = 0.
\end{split}
\end{equation*}
Combining this with \eqref{claim:p=0}, we get the claim \eqref{xxx:2} for
$p=0$. The proof of the induction step follows the same line as in the proof of
Lemma \ref{lem:4:6} and will, therefore, be omitted.

By the absolute convergence of the series \eqref{3:9:neu}, from
\eqref{xxx:2} it follows that
\begin{equation*}
\lim_{p\rightarrow\infty} \sum_{\bw\in F_{\fs, v_0}(p)} W_{\cM}(\bw)
\int_{I_{j(\bw)}} g_t(2x_{j(\bw)}+|\bw|) dx_{j(\bw)}=0,
\end{equation*}
which completes the proof of the lemma.
\end{proof}

To complete the proof of Theorem \ref{thm:4:1} it remains to consider the
particular case of magnetic perturbations of standard boundary conditions
(see Examples \ref{3:ex:3} and \ref{3:ex:mag}). We assume that the maximal
isotropic subspace $\cM$ corresponds to the magnetic perturbation of the
Laplace operator $-\Delta(\cM_{\mathrm{st}},\underline{a})$ with standard
boundary conditions, that is,
\begin{equation*}
\cM=\cM_{\mathrm{st}}^U\quad\text{with}\quad U=\bigoplus_{v\in V} U_v,\quad
U_v=\diag\left(\{\e^{\ii\varphi_j(v)}\}_{j\in\cS(v)}\right).
\end{equation*}

First we calculate $\tr_\cK \fS(\cM)$. Using \eqref{1:thm:gt},
\eqref{standard:ee}, and \eqref{spur} we get
\begin{equation*}
\begin{split}
\tr_\cK \fS(\cM) & = \tr_\cK \fS(\cM_{\mathrm{st}}) = \sum_{v\in V}
\sum_{j\in\cS(v)} [\fS_v(\cM_{\mathrm{st}})]_{j,j}\\
&= \sum_{v\in V} (2-\deg(v)) = 2|V|-|\cE|-2|\cI|.
\end{split}
\end{equation*}

Let $H_1(\cG,\Z)$ be the first homology group of the interior
$\cG_{\mathrm{int}}$ of the graph $\cG$. There is a canonical map $\Gamma:
\mathfrak{C}\to H_1(\cG,\Z)$, which satisfies
\begin{equation*}
\Gamma(\mathfrak{c}_{rev})=-\Gamma(\mathfrak{c}),\qquad
\Gamma(p\mathfrak{c})=p\Gamma(\mathfrak{c}),\qquad p\in\N.
\end{equation*}
In particular, $\Gamma(\mathfrak{c})=0$ when
$\mathfrak{c}=\mathfrak{c}_{\mathrm{rev}}$. Therefore, the map $\Gamma$ is
not injective. In general it is also not surjective. For any cycle
$\fc\in\fC$ we set
\begin{equation*}
\Phi(\fc):=\widetilde{\Phi}(\Gamma(\fc)),
\end{equation*}
where $\widetilde{\Phi}(\Gamma(\fc))$ is the magnetic flux through the
homological cycle $\Gamma(\fc)$ as defined in \cite{KS5}. If
\begin{equation*}
\bw=\{j_0,v_0,j_1,\ldots,j_n,v_n,j_0\}
\end{equation*}
is an arbitrary walk in the equivalence class $\fc$, then by explicit calculations the magnetic flux
through the cycle $\fc$ (see \cite{KS5}) is given by
\begin{equation}\label{def:flux}
\Phi(\fc) = \sum_{k=0}^{n-1} \left(\varphi_{j_k}(v_k) -
\varphi_{j_{k+1}}(v_k)\right) + \left(\varphi_{j_n}(v_n)-\varphi_{j_0}(v_n)\right).
\end{equation}
Obviously,
\begin{equation*}
W_{\cM}(\fc) = W_{\cM_{\mathrm{st}}}(\fc) \e^{\ii\Phi(\fc)}
\end{equation*}
and
\begin{equation}\label{flux:rev}
\Phi(\fc_{\mathrm{rev}})=-\Phi(\fc).
\end{equation}
By Proposition \ref{k-independent}, we have
$W_{\cM}(\fc_{\mathrm{rev}})=\overline{W_{\cM}(\fc)}$. In particular, if
$\fc_{\mathrm{rev}}=\fc$, then $W_{\cM}(\fc)$ is real. Since
$\Phi(\fc_{\mathrm{rev}})=-\Phi(\fc)$ and
$W_{\cM_{\mathrm{st}}}(\fc_{\mathrm{rev}})=W_{\cM_{\mathrm{st}}}(\fc)$, the
factor $\e^{\ii p\Phi(\fc)}$ in \eqref{gutzw} can be replaced by
$\cos(p\Phi(\fc))$. This completes the proof of Theorem \ref{thm:4:1}.

\section{Applications to Inverse Problems}\label{sec:inv:probl}

In this section we present an application of the trace formula in Theorem
\ref{thm:4:1} to inverse spectral and scattering problems. Throughout the
whole section we will assume that the maximal isotropic subspace $\cM$
satisfies any of the equivalent assumption of Proposition
\ref{k-independent}.

For the noncompact graph $\cG$ let $S(\lambda;\cM,\underline{a}):
\cK_{\cE}\rightarrow\cK_{\cE}$, $\lambda>0$, be the scattering matrix for
the triple $(-\Delta(\cM;\underline{a}),$  $-\Delta_+, \cJ)$ defined in
\cite{KS} according to the scattering theory in two Hilbert spaces
\cite[Chapter 2]{Yafaev}. Here $\cJ$ is the identification operator defined
in Section \ref{sec:trac:form}. The scattering matrix is continuous with
respect to the spectral parameter $\lambda>0$ (see \cite{KS1} or
\cite[Theorem 3.12]{KS4}).

Let $\xi(\lambda;\cM,\underline{a})$ be the spectral shift function
associated with the triple $(-\Delta(\cM;\underline{a}),$  $-\Delta_+,
\cJ)$ (see \cite[Section 8.11]{Yafaev}). It satisfies the trace formula
\begin{equation}\label{trace}
\begin{split}
\tr_{\cH}\left[\e^{t\Delta(\cM;\underline{a})}-\cJ\e^{t\Delta_+}\cJ^\dagger
\right] & + \tr_{\cH_{\cE}}\left[(\cJ^\dagger\cJ-I_{\cH_{\cE}})\e^{t\Delta_+}\right] \\
& = -t\int_0^\infty
\xi(\lambda;\cM,\underline{a})\e^{-t\lambda}d\lambda,\quad t>0,
\end{split}
\end{equation}
and is fixed uniquely by the condition $\xi(-1;\cM,\underline{a})=0$. From
the definition of the operator $\cJ$ it follows that the second term on the
r.h.s.\ of \eqref{trace} vanishes. Thus,
\begin{equation*}
\tr_{\cH}\left(\e^{t\Delta(\cM,\underline{a})} - \cJ \e^{t\Delta_+}
\cJ^\dagger \right) =  -t \int_0^\infty \e^{-\lambda t} \xi(\lambda;\cM)
d\lambda ,\quad t>0.
\end{equation*}
By the Birman-Krein theorem the spectral shift function is related to the
scattering matrix,
\begin{equation}\label{Birman:Krein}
\det_{\cK_{\cE}} S(\lambda;\cM, \underline{a})=\exp\{-2\pi i\xi(\lambda;\cM,
\underline{a})\}\qquad\text{a.e.}\quad\lambda\in\R_+.
\end{equation}

By the continuity of the scattering matrix and due to \eqref{Birman:Krein}
one can choose the branch of the logarithm such that
\begin{equation}\label{scatt:phase}
s(\lambda; \cM,\underline{a}):= \frac{1}{2\ii}\log\det_{\cK_{\cE}}
S(\lambda;\cM,\underline{a})
\end{equation}
is continuous with respect to $\lambda\in(0,\infty)$ and satisfies
\begin{equation*}
s(0+; \cM,\underline{a}) =-\pi\left(\xi(0+; \cM,\underline{a})+N(0+; \cM,\underline{a}) \right),
\end{equation*}
where $N(\lambda; \cM,\underline{a})$ is the counting function for the eigenvalues of the
operator $-\Delta(\cM,\underline{a})$. The function \eqref{scatt:phase} is
called the \emph{scattering phase}. The scattering phase and the eigenvalue
counting function uniquely determine  the spectral shift function,
\begin{equation*}
\xi(\lambda;\cM,\underline{a}) = -\frac{1}{\pi}
s(\lambda;\cM,\underline{a}) - N(\lambda; \cM,\underline{a}), \quad\lambda\in\R_+.
\end{equation*}

For compact graphs the spectral shift function is determined by the eigenvalue counting function alone,
$\xi(\lambda;\cM,\underline{a}) = - N(\lambda; \cM,\underline{a})$, $\lambda\in\R_+$.

\begin{proposition}\label{propo:5:1}
Assume that the graph $\cG$ (compact or noncompact) has no tadpoles. Let
the maximal isotropic subspace $\cM$ satisfy any of the equivalent
conditions of Proposition \ref{k-independent} and define local boundary
conditions. Then the spectral shift function $\R_+\ni\lambda\mapsto
\xi(\lambda)$ uniquely determines the set
\begin{equation}\label{red:spectrum}
\Bigl\{\ell>0\,\Big|\, \sum_{\substack{\fc\in\fC\\ |\fc|=\ell}} W_{\cM}(\fc)\neq 0 \Bigr\}.
\end{equation}
\end{proposition}

In \cite{KuNo} the set \eqref{red:spectrum} is called  the ``reduced length
spectrum''.

\begin{proof}
Using standard formulas for the inverse Laplace transform and, in
particular, the fact that $t^{-3/2} \e^{-a/t}$, $a>0$, is the Laplace transform of
$(\pi a)^{-1/2}\sin(2\sqrt{a\lambda})$, from Theorem \ref{thm:4:1}
we get
\begin{equation}\label{xi:per}
\begin{split}
-\xi(\lambda;\cM,\underline{a}) & = \frac{L}{\pi}\sqrt{\lambda} +
\frac{1}{4}\tr_{\cK} \fS(\cM)-\frac{|\cE|}{4}\\ & + \frac{1}{\pi}
\sum_{\fc\in\fC_{\mathrm{prim}}} \sum_{p\in\N} \frac{1}{p} W_{\cM}(\fc)^p
\sin\left(p|\fc| \sqrt{\lambda}\right),\quad \lambda > 0,
\end{split}
\end{equation}
where the series converges in the sense of distributions on
$\R_+$. For $\sk\in\R$ define the function $u(\sk)$ via
\begin{equation*}
u(\sk) := \begin{cases} -\xi(\sk^2; \cM,\underline{a}) & \text{if}\quad
\sk>0,\\ \phantom{-}\xi(\sk^2; \cM,\underline{a}) & \text{if}\quad \sk<0.
\end{cases}
\end{equation*}
Using \eqref{xi:per} we can calculate the distributional derivative of $u$,
\begin{equation*}
\begin{split}
u'(\sk) & = \frac{L}{\pi} + \left(\frac{1}{2}\tr_{\cK}
\fS(\cM)-\frac{|\cE|}{2}\right) \delta(\sk)\\ & + \frac{1}{\pi}
\sum_{\fc\in\fC_{\mathrm{prim}}} \sum_{p\in\N} W_{\cM}(\fc)^p |\fc|
\cos\left(p|\fc| \sk\right),\quad \sk\in\R,
\end{split}
\end{equation*}
where $\delta$ stands for the Dirac $\delta$-distribution.
Its Fourier transform with respect to $\sk$ yields
\begin{equation*}
\begin{split}
\int_\R \e^{\ii\omega\sk} u'(\sk) d\sk & =2L \delta(\omega) +
\frac{1}{2}\tr_{\cK} \fS(\cM)-\frac{|\cE|}{2}\\ & +
\sum_{\fc\in\fC_{\mathrm{prim}}} \sum_{p\in\N} W_{\cM}(\fc)^p |\fc|
\left[\delta(\omega-p|\fc|) + \delta(\omega+p|\fc|)\right],
\end{split}
\end{equation*}
which implies the claim.
\end{proof}

\begin{theorem}\label{thm:inv:probl}
Assume that the graph $\cG$ (compact or noncompact) has no tadpoles. Let
the maximal isotropic subspace $\cM$ satisfy any of the equivalent
conditions of Proposition \ref{k-independent} and define local boundary
conditions on the graph $\cG$. Assume, in addition, that
\begin{itemize}
\item[(i)]{the lengths $a_{i}\,(i\in\cI)$ of the internal edges of the
graph $\cG$ are rationally independent, that is, the equation
\begin{equation*}
\sum_{i\in\cI}n_{i}\, a_{i} = 0
\end{equation*}
with \emph{integer} $n_{i}\in\Z$ has no non-trivial solution;}
\item[(ii)]{for any vertex $v\in V$, none of the matrix elements $[\fS_v]_{j,j^\prime}$,
$j,j^\prime\in\cS(v)$, vanishes.}
\end{itemize}
Then the spectral shift function $\R_+\ni\lambda\mapsto \xi(\lambda)$
uniquely determines the interior $\cG_{\mathrm{int}}$ of the graph $\cG$.
\end{theorem}

\begin{proof}
All arguments of Section 4 in \cite{KuNo} remain valid for boundary
conditions satisfying assumption (ii) of the theorem. Thus, the set
\eqref{red:spectrum} uniquely determines the interior $\cG_{\mathrm{int}}$
of the graph $\cG$. Combining this with Proposition \ref{propo:5:1} we
obtain the claim.
\end{proof}

Note that assumption (ii) of Theorem \ref{thm:inv:probl} implies that if
$\cM_v$ corresponds to the standard boundary conditions at the vertex $v$
or its magnetic perturbation, then necessarily $\deg(v)\neq 2$.

\begin{remark}
As in \cite{Novaszek} the assumption on the rational independence of the
edge lengths can be slightly relaxed.
\end{remark}


\end{document}